\documentclass[10pt,conference]{IEEEtran}
\IEEEoverridecommandlockouts

\usepackage[utf8]{inputenc}
\usepackage{amsmath,amsfonts}
\usepackage{amsmath, amsthm, amssymb}

\ifcase 2  %% select the font to use
\or
\usepackage{inconsolata}
\or
\fi

\theoremstyle{definition}

\theoremstyle{plain}

\theoremstyle{remark}

\usepackage{subcaption}

\usepackage{algorithmic}
\usepackage{graphicx}
\usepackage{textcomp}
\usepackage[inline]{enumitem}
\usepackage{url}

\usepackage[T1]{fontenc}
\usepackage[utf8]{inputenc}
\usepackage{nicefrac}
\usepackage{siunitx}  %% formatting numbers and units
\usepackage{flushend}
\usepackage{hyperref}

\usepackage{booktabs} 
\SetEnumitemKey{midsep}{topsep=3pt, partopsep=0pt}
\usepackage{todonotes}
\newif\iftodos
\todostrue
\iftodos
\newcommand{\PP}[1]{\todo[inline]{#1}}
\newcommand{\JN}[1]{\todo[inline,color=green!40]{JN: #1}}
\newcommand{\jn}[1]{\todo[color=green!40,size=\tiny]{JN: #1}}
\newcommand{\MF}[1]{\todo[inline,color=blue!40]{MF: #1}}
\else
\newcommand{\PP}[1]{}
\newcommand{\JN}[1]{}
\newcommand{\jn}[1]{}
\newcommand{\MF}[1]{}
\fi

\def\BibTeX{{\rm B\kern-.05em{\sc i\kern-.025em b}\kern-.08em
    T\kern-.1667em\lower.7ex\hbox{E}\kern-.125emX}}
\begin{document}

\title{Out of Sight, Still at Risk: The Lifecycle of Transitive Vulnerabilities in Maven}

\author{
  \IEEEauthorblockN{%%
    Piotr Przymus\IEEEauthorrefmark{1},
    Mikołaj Fejzer\IEEEauthorrefmark{2},
    Jakub Narębski\IEEEauthorrefmark{3},
    Krzysztof Rykaczewski\IEEEauthorrefmark{4} and
    Krzysztof Stencel\IEEEauthorrefmark{5}
  }

  \textit{Nicolaus Copernicus University in Toruń},\quad
  \IEEEauthorrefmark{5}\textit{University of Warsaw}\\
  \null\hspace*{1.9cm} Toruń, Poland, \hspace{3cm} Warsaw, Poland,
  \\
  Email:
  \IEEEauthorrefmark{1}piotr.przymus, \IEEEauthorrefmark{2}mfejzer, \IEEEauthorrefmark{3}jakub.narebski, \IEEEauthorrefmark{4}krzysztof.rykaczewski@mat.umk.pl\\
  \IEEEauthorrefmark{5}stencel@mimuw.edu.pl
}

\IEEEtitleabstractindextext{
\begin{abstract}
The modern software development landscape heavily relies on transitive dependencies.
They enable seamless integration of third-party libraries.
However, they also introduce security challenges.
Transitive vulnerabilities that arise from indirect dependencies expose projects to risks associated with Common Vulnerabilities and Exposures (CVEs).
It happens even when direct dependencies remain secure.

This paper examines the lifecycle of transitive vulnerabilities in the Maven ecosystem.
We employ survival analysis to measure the time projects remain exposed after a CVE is introduced.
Using a large dataset of Maven projects, we identify factors that influence the resolution of these vulnerabilities.
Our findings offer practical advice on improving dependency management.
\end{abstract}

}

\maketitle
\IEEEdisplaynontitleabstractindextext

\definecolor{mygray}{RGB}{240,240,240}

\newcommand{\answer}[2]{\vspace{.2cm}{\centering\setlength{\fboxrule}{0.1pt}\fbox{\colorbox{mygray}{\parbox{0.95\columnwidth}{\textbf{Answer to RQ}. #2}}}\vspace{.2cm}}}

% page limit = 4 + references

\maketitle

\makeatletter
\newcommand\footnoteref[1]{\protected@xdef\@thefnmark{\ref{#1}}\@footnotemark}
\makeatother

\section{Introduction}\label{sec:introduction}

Modern software development relies on third-party libraries to accelerate progress and enhance functionality. 
However, such libraries may introduce transitive dependencies, i.e., indirect dependencies automatically included within the dependency graph. 
While transitive dependencies streamline workflows, they also present hidden risks~\cite{DBLP:journals/dtrap/DusingH22}, particularly security vulnerabilities. 
A single vulnerable library in the dependency graph can compromise an entire project through CVEs~\cite{cve-db}.
It is especially significant in ecosystems like Maven, where deeply nested and complex dependency graphs are prevalent.

The MSR 2025 Challenge~\cite{msr2025} provides an opportunity to investigate these challenges by analyzing dependencies within the Maven Central ecosystem using the Goblin framework. 
Goblin~\cite{goblinJaimeHP24} combines a Neo4J-based dependency graph~\cite{jaime_2024_13734581} with Weaver, i.e., a tool for customizable metric computation.
It facilitates research on complex software ecosystems.

This study investigates how the depth of a vulnerable dependency in a project's transitive dependency graph impacts the duration of the project's exposure to this vulnerability. 
We aim to answer the following research question:

\textbf{RQ}: How does the depth of transitive dependencies influence the time to fix vulnerabilities?

We apply survival analysis to evaluate the duration for which projects remain vulnerable after the introduction of a CVE.
Additionally, we hypothesise a model to represent this behavior.
% Replication package for this research is available at Figshare (analysed and extracted data from Goblin) ~\cite{https://figshare.com/s/2dde74dabf1628c69132} and on \PP{link do repo} (source code, and notebooks).
%% NOTE: this version has nicer layout, and doesn't take more space
The replication package, including extracted data and code, is available on Figshare~\url{https://doi.org/10.6084/m9.figshare.27956667}.

\section{Preliminaries}%
\label{sec:Preliminaries}

\textbf{Survival analysis}~\cite{liu2012survival} is a tool for analyzing time-to-event data, such as the failure of a mechanical component or the resolution of a vulnerability.
Let a random variable represent the time of the event of interest. 
The \emph{survival function}, $S(t)$, defines the probability that the event has not occurred by time $t$. 
If $F(t)$ is the cumulative distribution function of this random variable, the survival function is given by: $S(t) = 1 - F(t)$.

We compute the survival function using the \emph{Kaplan-Meier} estimator~\cite{doi:10.1080/01621459.1958.10501452}.
Let $d_i$ be the number of events occurring at time $t_i$, and $n_i$ the number of individuals who survived just prior to $t_i$.
The estimate of the survival function is:
$\hat{S}(t) = \prod_{i \mid t_i \leq t} ( 1 - \tfrac{d_i}{n_i} )$.

\section{Methods}\label{methods}
We use survival analysis with the Kaplan-Meier estimator to track how long projects remain vulnerable after CVE is introduced in a transitive dependency. 
Statistical modeling and regression assess the impact of transitive dependencies on fix delays. 
Below, we define some key notions.

\textbf{CVE Lifetime}
is the duration a software artifact remains exposed to a transitive vulnerability. 
We track whether a project's transitive dependencies include versions affected by a CVE, analyzing each vulnerability separately. 
If an artifact is impacted by multiple CVEs, each is processed independently. 
An artifact is no longer vulnerable when a version removes the dependency path to the vulnerable component or updates it to a non-affected version.

\textbf{Next Release} is computed for each artifact as the version following the current one, determined using Semantic Versioning~\cite{preston-wernerSemanticVersioning200}.
We cross-validate this selection with the following heuristic:  
(1) Check for a newer minor version.\,  
(2) If none exists, increment the penultimate version part and reset the minor and verify if the new selection is valid.\,  
(3) As a fallback, choose the oldest available newer version.  

Our analysis shows that in \textbf{95\% of cases}, the next version selected using Semantic Versioning matches the heuristic. 
To address non-standard and inconsistent versioning schemes, we allowed heuristic to cover remaining cases, increasing coverage to \textbf{96\%}. Additionally, we established edges linking each version to its successor, enabling sequential update tracking.

\textbf{Affected Versions} is a meta information specifying that this version is affected.
We start by extending the graph with CVE-related information, supplemented by data from NVD~\cite{nvd}. 
We start with identified all artifact versions directly affected by CVEs.
Then, we propagated this information to projects depending on these versions, identifying reverse transitive dependencies.
Propagating CVEs across the entire graph posed two key challenges. 
(1) Including all affected versions could introduce noise, as CVEs often span multiple versions over months or even years. 
(2) It introduces a significant computational overhead.
To address these challenges, we focused on the most recent affected version of each artifact (i.e.\ the version with the highest version number).
This ensured that the subsequent version, calculated earlier, was unaffected (as this was the youngest affected version, so next release is unaffected). 
The same approach was used to propagate transitive vulnerability information across dependency levels.
It enhanced computational efficiency and reduced informational noise while maintaining accuracy.

\textbf{Dataset:}
The challenge dataset graph~\cite{msr2025,jaime_2024_13734581} was extended with new edge types: \texttt{Mvn\_dep}, \texttt{NextRelease}, and \texttt{Affected}.
\texttt{Mvn\_dep} edges connect \texttt{Release} nodes.
They represent dependencies between project releases as defined in Maven's \texttt{pom.xml}.
They were precomputed based on graph structure.
\texttt{Affected} edges link \texttt{CVE} nodes to \texttt{Release} nodes.
They indicate affected versions.
A \texttt{NextRelease} edge points the next computed release for each artifact.

\section{Results}\label{results}

\subsection{Data characteristics}%
\label{sub:data_characteristics}

Our analysis includes over \num{132000} projects, encompassing more than 3 million artifacts and \num{2676} identified CVEs, spanning 19 years of development (see~Tab.~\ref{tab:characteristics}).
We used all artifacts present in the Neo4J database~\cite{msr2025,jaime_2024_13734581} cross referenced with Maven subset of Open source vulnerability DB~\cite{osv}.
The scale and diversity of the selected data provides a strong foundation for robust statistical and survival analyses.

We had to address inaccuracies in computation of the next version.
Thus, we removed all propagated nodes where the transition from a previous version to the next resulted in negative time intervals.
Similarly, nodes were removed if the time from the appearance of a CVE to its resolution was negative.
This step ensured the consistency and reliability of the dataset.
See Tab.~\ref{tab:characteristics} for how filtering affected the dataset.

\begin{table}[htb]
    \centering
    \caption{Dataset characteristics,  counts denoted as \#.}\label{tab:characteristics}
    \begin{tabular}{@{}lrrrc@{}}
      \toprule
                    & \# CVE     & \# Projects  & \# Assets     & Date Range \\
      \midrule
      All Data      & \num{2853} & \num{211369} & \num{4767177} & 2005-08 -- 2024-08 \\
      Filtered Data & \num{2676} & \num{132235} & \num{3383100} & 2005-08 -- 2024-08 \\
      \bottomrule
    \end{tabular}
\end{table}

\subsection{Survival analysis}%
\label{sub:Survival analysis}

\begin{figure}[htpb]
    \centering
    \includegraphics[width=0.5\textwidth]{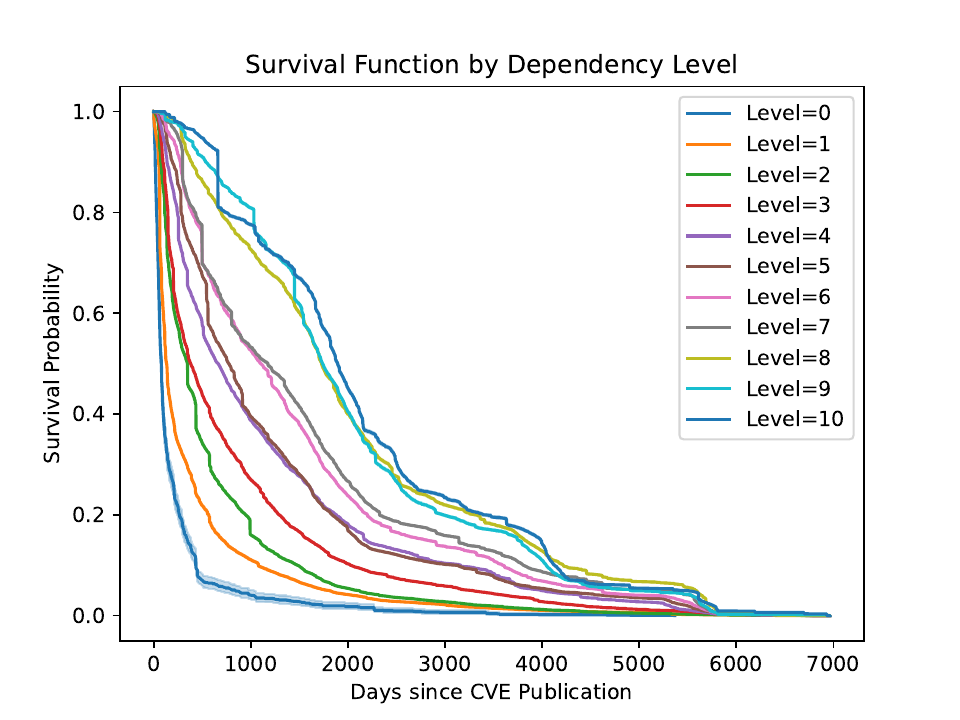}
    \caption{CVEs survival by dependency level.}
    \label{fig:surv_anal}
\end{figure}

To address \textbf{RQ} we used the survival analysis and the Kaplan-Meier estimator.
Fig.~\ref{fig:surv_anal} reveals a significant impact of transitive vulnerability depths on CVE persistence.
The survival function stratified by dependency levels shows that vulnerabilities at deeper levels of the dependency graph tend to persist longer compared to those at shallower levels.

\begin{table*}
  \centering
  \caption{Cumulative time to fix (from CVE to fix) and single-level time to fix (from faulty version to fix).}
  \label{tab:Survival analysis}
  \begin{tabular}{lrrrrrrrrrrrrrrr}
    \toprule
    & \multicolumn{7}{c}{Cumulative survival} & \multicolumn{7}{c}{Level survival} &\\
    \addlinespace[-0.5\defaultaddspace]
    \cmidrule(lr){2-8} \cmidrule(lr){9-15}
    \addlinespace[-0.4\defaultaddspace]
    level & mean    & std      & min & 25\%&       50\% &       75\% & max       & mean & std & min &25\%&50\%& 75\%& max        &   count\\
    \midrule
   \ 0 & \num{ 215} & \num{ 475} & 0 & 28  & \num{  65} & \num{ 175} & \num{5367} & 146 & 288 &   0 & 27 & 62 & 144 & \num{4663} & \num{3046} \\
   \ 1 & \num{ 429} & \num{ 759} & 0 & 62  & \num{ 140} & \num{ 450} & \num{6954} &  93 & 184 &   0 & 23 & 40 &  84 & \num{4101} & \num{184579} \\
   \ 2 & \num{ 597} & \num{ 824} & 0 & 133 & \num{ 334} & \num{ 699} & \num{6960} &  87 & 159 &   0 & 19 & 40 &  93 & \num{4060} & \num{417797} \\
   \ 3 & \num{ 792} & \num{1055} & 0 & 146 & \num{ 355} & \num{ 971} & \num{6967} &  83 & 145 &   0 & 21 & 41 &  90 & \num{3200} & \num{439878} \\
   \ 4 & \num{1104} & \num{1256} & 0 & 255 & \num{ 584} & \num{1534} & \num{6960} &  82 & 133 &   0 & 21 & 44 &  93 & \num{3055} & \num{455303} \\
   \ 5 & \num{1145} & \num{1240} & 0 & 296 & \num{ 676} & \num{1541} & \num{6958} &  78 & 125 &   0 & 19 & 39 &  85 & \num{4202} & \num{482524} \\
   \ 6 & \num{1364} & \num{1333} & 1 & 381 & \num{ 888} & \num{1846} & \num{6964} &  74 & 113 &   0 & 21 & 41 &  81 & \num{3114} & \num{428033} \\
   \ 7 & \num{1492} & \num{1411} & 1 & 474 & \num{ 958} & \num{1990} & \num{6964} &  72 & 119 &   0 & 25 & 40 &  69 & \num{3055} & \num{375772} \\
   \ 8 & \num{1900} & \num{1485} & 1 & 692 & \num{1617} & \num{2522} & \num{6964} &  73 & 128 &   0 & 20 & 37 &  76 & \num{3055} & \num{248152} \\
   \ 9 & \num{1943} & \num{1359} & 3 & 989 & \num{1642} & \num{2503} & \num{6964} &  68 & 114 &   0 & 22 & 38 &  75 & \num{3055} & \num{198331} \\
    10 & \num{2075} & \num{1426} & 0 & 874 & \num{1822} & \num{2673} & \num{6965} &  59 &  87 &   0 & 21 & 41 &  62 & \num{2074} & \num{149685} \\
\bottomrule
\end{tabular}
\end{table*}

The analysis of moments in cumulative and single-level time to fix is summarized in Tab.~\ref{tab:Survival analysis}.
It reveals distinct trends regarding the impact of dependency depths on vulnerability resolutions.
The cumulative time to fix measured from a CVE introduction to its resolution increases with dependency depth.
The mean times rise steadily from \num{215} days at level 0 to \num{2075} days at level 10. 
Similarly, median cumulative times also increase.
It suggests a compounding delay effect as vulnerabilities propagate through deeper levels of the dependency graph. 
The standard deviation is higher at greater depths.
It indicates diverse timeframes for resolving vulnerabilities in complex deeply nested dependencies.

Conversely, the single-level time to fix, i.e.\ the time needed to deliver the next release of an artifact that is CVE free, remains relatively consistent across levels.
Mean times decrease slightly from \num{146} days at level 0 to \num{59} days at level 10.
Median times start stabilizing from level 1.
To summarise delays are predominantly due to the cumulative propagation rather than inefficiencies within individual levels.

We also observe that the majority of transitive dependencies are concentrated mid-level (see the last column of Tab.~\ref{tab:Survival analysis}).
The pick value of \num{482524} dependencies occurs at the intermediate level of \num{5}.  
Thus, the cumulative repair delays at such levels are further amplified.

\subsubsection*{Distribution Fitting of Resolution Times}

\begin{figure}[ht]
    \centering
    \centering
    \includegraphics[width=0.9\linewidth]{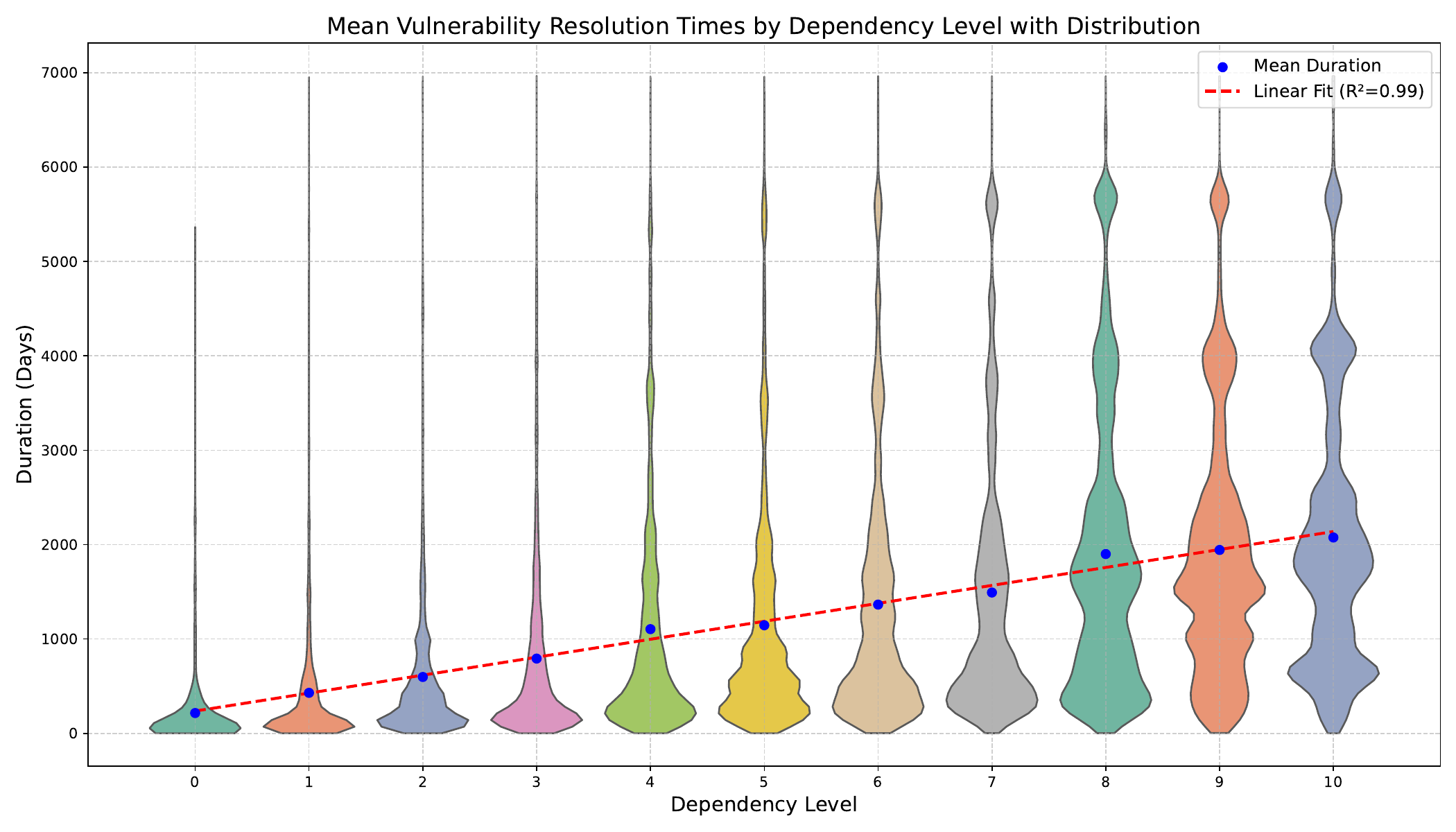}
    \caption{Mean resolution times across dependency levels with fitted linear regression model.}
    \label{fig:mean_violin_regression_plot}
\end{figure}

\begin{table}[ht]
    \centering
    \caption{Goodness-of-fit statistics for various distributions applied to vulnerability resolution times.}
    \begin{tabular}{lrrr}
    \toprule
    \multicolumn{1}{c}{\textbf{Distribution}} &  \multicolumn{1}{c}{\textbf{AIC}} & \multicolumn{1}{c}{\textbf{A-D}} & \multicolumn{1}{c}{\textbf{p-value A-D}}  \\
    \midrule
    Exponential &  \num{54692119} & \num{33598} & 0.01\\
    Weibull     &  \num{54658206} & \num{10327} & 0.01\\
    Gamma       &  \num{54676357} & \num{   59} & 1\\
    Log-Normal  &  \num{54720321} & \num{   35} & 1\\
    \bottomrule
    \end{tabular}
    \label{tab:distribution_fitting}
\end{table}

To elaborate further on \textbf{RQ}, we examined the underlying distribution of vulnerability resolution times across various dependency levels.

A visual inspections of the violin plot (Fig.~\ref{fig:mean_violin_regression_plot}) revealed that the data exhibited long tails.
Thus, the distribution seemed to be skewed rather than symmetric.
This observation led us to hypothesize that the resolution times might follow a Gamma distribution.
To assess the suitability of different distributions, we fitted Exponential, Weibull, Gamma, and Log-Normal distributions to the empirical data.
We evaluated their goodness-of-fit using the Akaike Information Criterion (AIC) and Anderson-Darling (A-D) test statistics.
The summary of the fitting results is presented in Tab.~\ref{tab:distribution_fitting}.

The Exponential distribution had the highest A-D statistics and lowest p-value, leading to its rejection.
Weibull performed better but was still rejected (p = 0.01).
Both Gamma and Log-Normal passed (p = 1), but Gamma is better suited for modeling failure times.
While Log-Normal fits lower dependency levels, Q-Q plots show deviations at higher levels, whereas Gamma fits well across all levels (see replication package).

\subsection{Linear Regression Analysis of Mean and Median}\label{subsec:LinearRegression}

\begin{figure}[ht]
    \centering
    \includegraphics[trim={0 0 0 0.5cm},clip, width=0.8\linewidth]{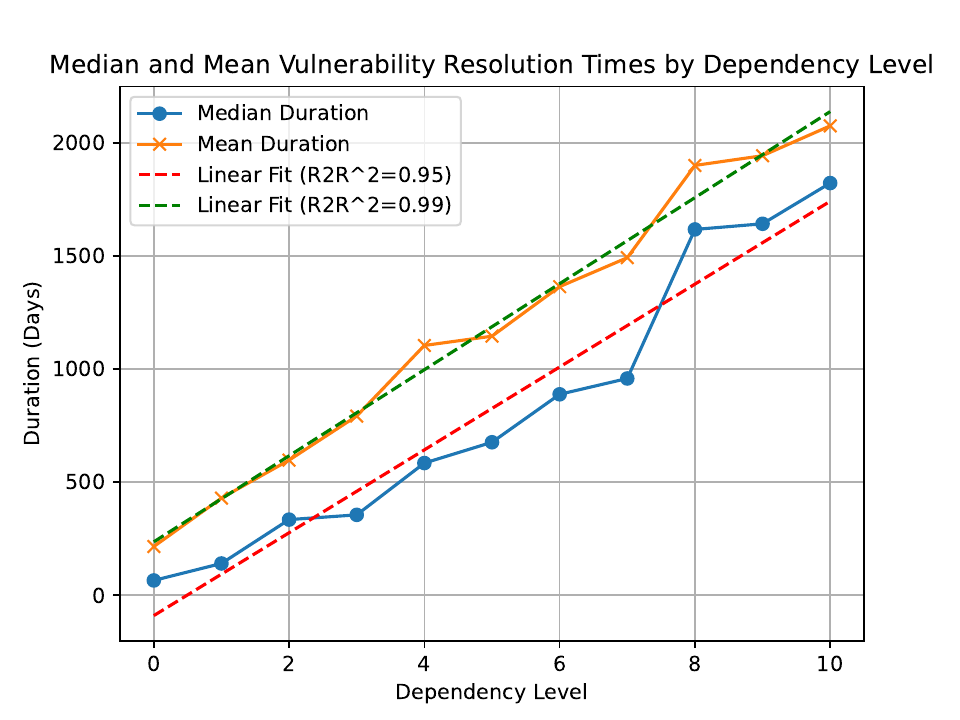}
    \caption{Linear regression analysis of mean and median vulnerability resolution times.}
    \label{fig:linear_fit}
\end{figure}

To finally answer \textbf{RQ} we performed linear regression analyses on both the median and mean repair durations across different dependency levels. 

For the median resolution times, the linear model is given by
$ \text{Median Duration} = -90.14 + 183.15 \times \text{Level}. $
This model achieves an $R^2$ value of 0.9466.
It indicates that approximately 94.66\% of the variability in median resolution times can be explained by the dependency level.

Similarly, for the mean resolution times, the linear model is expressed as
$\text{Mean Duration} = 238.05 + 189.92 \times \text{Level}.$
The corresponding $R^2$ value is 0.9890.
It suggests an even stronger explanatory power, since 98.90\% of the variance in mean resolution times is attributable to the dependency level.

These findings confirm that higher dependency levels are strongly associated with longer vulnerability resolution times, supporting our initial hypothesis that deeper dependencies lead to extended exposure durations.

\section{Discussion}\label{discussion}

\answer{}{
Each additional level of transitive dependency increases CVE resolution time and extends a project's vulnerability.
It should be considered when assessing risks associated with projects.
Approximately: $\textbf{Mean CVE lifetime} \approx (\text{Level} \times 6 \text{ months}) + 8 \text{ months}, $
$\textbf{Median CVE lifetime} \approx (\text{Level} \times 6 \text{ months}) - 3 \text{ months}.$
}

Linear regression analyses presented in Section~\ref{subsec:LinearRegression} reveal a robust positive correlation between dependency levels and both median and mean vulnerability resolution times.
High $R^2$ values indicate that dependency levels are significant predictors of how long vulnerabilities persist within projects.

We hypothesize that a graph-based mathematical model can clarify the relationship between dependency levels and vulnerability resolution times. 
For that we propose a mathematical model to explain the observed data phenomena.
We represent dependencies between revisions as a directed acyclic graph $G = (V, E)$, where each node $v \in V$ denotes a software component, and each edge $(u \to v) \in E$ signifies that $v$ depends on $u$.
When a vulnerability is detected and fixed in a base component $v_0$, the fix propagates to all dependent components directly or indirectly.

The \textbf{resolution rate} $\beta(u)$ for a vulnerability in library $u$ is an inverse function of its dependency depth $\beta(u) = \frac{k}{d(u) + 1}$, where $k > 0$ is a constant and $d(u)$ is the dependency depth.

Assume the resolution process has $\alpha$ independent stages, each with time $X_i$ that follows Gamma distribution.
Then the resolution time $T_u = \sum_{i=1}^\alpha X_i \sim \text{Gamma}(\alpha, \beta(u))$.
This shows that the total resolution time follows a Gamma distribution under independent sequential stages.

In the graph-based model, the \textbf{expected resolution time} $\mathbb{E}[T_u]$ of a vulnerability in library $u$ is \textbf{linearly dependent} on its \textbf{dependency depth} $d(u)$:
$\mathbb{E}[T_u] = \alpha \frac{d(u) + c}{k}$.
We plan to investigate this hypothesized model in future work.

\section{Threats to Validity}

(1) \textbf{Issues with Identifying Next Versions}: We compared the time of artifact release between the current and calculated new version. Problematic observations such as previous minor version released with a later date than the next version of the same artifact were removed to mitigate inaccuracies.
\\\noindent
(2) \textbf{Focus on Youngest Vulnerable Versions}: Observing only the youngest fixes limits the scope but provides a reliable lower bound for estimating fix times in transitive dependencies.  
\\\noindent
(3) \textbf{Exclusion of Embargo Phase}: Fix times are measured from CVE publication, ignoring embargo periods or unreleased fixes, which may underestimate resolution times.
\\\noindent
(4) \textbf{Assumption of Vulnerability}: Transitive vulnerabilities do not always make a project vulnerable but increase the attack surface. For simplicity, we assume all projects with vulnerable dependencies are vulnerable.
Despite these limitations, the study offers valuable insights into transitive vulnerability dynamics in Maven.

\section{Related work}\label{related}
Survival analysis was applied to lifecycles of vulnerabilities~\cite{DBLP:journals/tse/IannoneGFLP23,DBLP:conf/msr/PrzymusFNS23} and dependencies~\cite{DBLP:journals/ese/PranaSSFSSL21} by numerous researchers in empirical studies discussed below.
Additionally various studies analyzed Maven ecosystem in terms of dependencies and upgrade cycle~\cite{DBLP:journals/ese/KulaGOII18,DBLP:journals/dtrap/DusingH22,DBLP:journals/ese/Soto-ValeroHMB21,DBLP:conf/kbse/ZhangLCXFZZL23}.
%To our knowledge, no prior studies have examined depth of dependency along with CVE classification.

\begin{enumerate*}
\item %%
Iannone et al.~\cite{DBLP:journals/tse/IannoneGFLP23} analyzed \num{1096} GitHub projects connected to \num{3663} NVD vulnerabilities to determine how long each vulnerability survives.
They used the SZZ algorithm and the Kaplan-Meier estimator.
They found out that at least half of the vulnerabilities survive until 511 days; a median of 9 changes is required are required to fix them; and developers are not aware of already present problems, lacking automated detection tools.

\item %%
Prana et al.~\cite{DBLP:journals/ese/PranaSSFSSL21} used Veracode SCA tool to detect dependencies for Java, Python and Ruby open source projects (\num{450} total).
They also used the Kaplan-Meier estimator. 
They ranked programming languages by the speed of their upgrade cycles, with Python being the slowest, Ruby the fastest, and Java falling in the middle.

\item %%
Przymus et al.~\cite{DBLP:conf/msr/PrzymusFNS23} investigated CVE lifetime per project using Kaplan-Meier estimator, according to various risk factors, such as CVE characteristics, programming language characteristics, project characteristics.
The dataset comprised \num{22700} CVEs cross referenced with commits from \num{9800} projects.
Project data was obtained via World of Code infrastructure~\cite{DBLP:journals/ese/MaDBAVTKZM21,msr-challenge2023}.
The most important factor were found to be the programming language memory model, the CVE attack vector and the number of project contributors.
Overall 75\% of fixes required between 1 to 3 commits for the fix, with the median fix time of 34 days.

\item %%
Kula et al.~\cite{DBLP:journals/ese/KulaGOII18} analyzed Maven \verb|pom.xml| files of \num{4659} GitHub projects and conducted a developer survey, to find that 81.5\% of analyzed projects still utilize outdated dependencies.
Required updates are seen by contributing developers as boring, low priority tasks, done in spare time.

\item %%
Düsing et al.~\cite{DBLP:journals/dtrap/DusingH22} analyzed dependency upgrades in Maven Central, NuGet.org, and the NPM Registry, comprising 1.9 million libraries and \num{3736} CVEs.
Only 1\% of libraries in NuGet have vulnerable dependencies while for Maven Central at least 29\% is similarly affected.
Upgrades of vulnerable dependencies usually happen more than 200 days prior to vulnerability publication.

\item %%
Soto-Valero et al.~\cite{DBLP:journals/ese/Soto-ValeroHMB21} investigated \num{723444} Maven dependency relations from \num{9639} programs, to detect not needed ones. 
As far as 57\% of transitive dependencies are bloated, compared to 2\% of direct dependencies. 

\item %%
Zhang et al.~\cite{DBLP:conf/kbse/ZhangLCXFZZL23} conducted a study to examine the prevalence of persistent vulnerabilities in the Maven ecosystem, using \num{1861} CVE cross referenced with \num{541753} libraries.
The authors found that 58.73\% of vulnerabilities were left in 50\% of affected projects and introduced Ranger, a tool suggesting updates along still compatible library ranges.

\end{enumerate*}

In comparison, our study provides unique insight into CVE vulnerability survival per Maven dependency level, which has not been previously researched.

\section{Conclusion}
This study investigates the survival of transitive vulnerabilities in the Maven ecosystem.
It emphasizes the substantial influence of the dependency depth on the time required to resolve vulnerabilities.
Our analysis shows that vulnerabilities at deeper levels persist longer due to compounded delays in transitive dependency resolution, even as single-level resolution times remain stable.

Our findings underscore the importance of managing transitive dependencies effectively, especially in projects developed with security in mind.
The dependency depth should be considered a critical factor in assessing project risk, as deeper vulnerabilities introduce considerable overhead to security fixes.
Future work will focus on developing a formalized model, evaluating it across different ecosystems, and exploring automated solutions to address the persistence of transitive vulnerabilities.

\newpage

\bibliographystyle{IEEEtran}
\bibliography{paper.bib}

\end{document}